\documentclass[twocolumn, prb, showpacs]{revtex4}
\usepackage{graphicx}
\usepackage{bm}
\usepackage{amssymb}
\usepackage{amsmath}
\newcommand{\be}{\begin{equation}}
\newcommand{\ee}{\end{equation}}
\newcommand{\bea}{\begin{eqnarray}}
\newcommand{\eea}{\end{eqnarray}}
\newcommand{\Ket}[1]{\vert \, #1 \, \rangle}

\newcommand{\MatEl}[3]{\langle \, #1 \,\vert\,#2\,\vert\,#3\,\rangle}
\newcommand{\Amp}[2]{\langle \, #1\, \vert\,  #2 \, \rangle}

\renewcommand{\phi}{\varphi}
\renewcommand{\epsilon}{\varepsilon}
\renewcommand{\vec}[1]{{\bm #1}}

\usepackage{color}

\begin{document}

\title{Spin relaxation due to deflection coupling in nanotube quantum dots}
\author{Mark S. Rudner$^1$ and Emmanuel I. Rashba$^{1, 2, 3}$}
\affiliation{
$^{(1)}$ 
 Department of Physics,
 Harvard University, 
 Cambridge, MA 02138\\
$^{(2)}$
 Center for Nanoscale Systems, 
 Harvard University,
 Cambridge, MA 02138\\
$^{(3)}$
 Department of Physics,
 Loughborough University,
 Leicestershire LE11 3TU, UK
}

\begin{abstract}
We consider relaxation of a  single electron spin in a nanotube quantum dot due to its coupling to flexural phonon modes, and identify a new spin-orbit mediated coupling between the nanotube deflection and the electron spin.
This mechanism dominates other spin relaxation mechanisms in the limit of small energy transfers.
Due to the quadratic dispersion law of long wavelength flexons, $\omega \propto q^2$, the density of states $dq/d\omega \propto \omega^{-1/2}$ diverges as $\omega \rightarrow 0$.
Furthermore, because here the spin couples directly to the nanotube deflection, there is an additional enhancement by a factor of $1/q$ compared to the deformation potential coupling mechanism.
We show that the deflection coupling robustly gives rise to a minimum in the magnetic field dependence of the spin lifetime $T_1$ near an avoided crossing between spin-orbit split levels in both the high and low-temperature limits.
This provides a mechanism that supports the identification of the observed $T_1$ minimum with an avoided crossing in the single particle spectrum by Churchill et al.[Phys. Rev. Lett. {\bf 102}, 166802 (2009)].

\end{abstract}

\maketitle

\section{Introduction} 
Due to their outstanding mechanical properties and versatile electrical characteristics, carbon nanotubes offer an exciting platform both for studies of fundamental physical phenomena and for a variety of potential applications.
The relatively small nuclear charge of carbon and the low natural abundance of carbon isotopes with nonzero nuclear spin suggest that the spin-orbit and hyperfine interactions, which are the main sources of electron spin relaxation in GaAs\cite{Petta, Hanson}, should be weak in carbon nanotubes.
Thus in recent years the electronic spin properties of nanotubes have gained wide attention for potential applications in spintronics and quantum computing.
Furthermore, the availability of isotopically purified starting materials opens the possibility of growing $^{12}$C (nuclear spin $I = 0$) and $^{13}$C (nuclear spin $I = 1/2$) nanotubes to study the behavior of electron spins in the presence or absence of a nuclear spin bath\cite{HughNP}.

\begin{figure}
\includegraphics[width=3.2in]{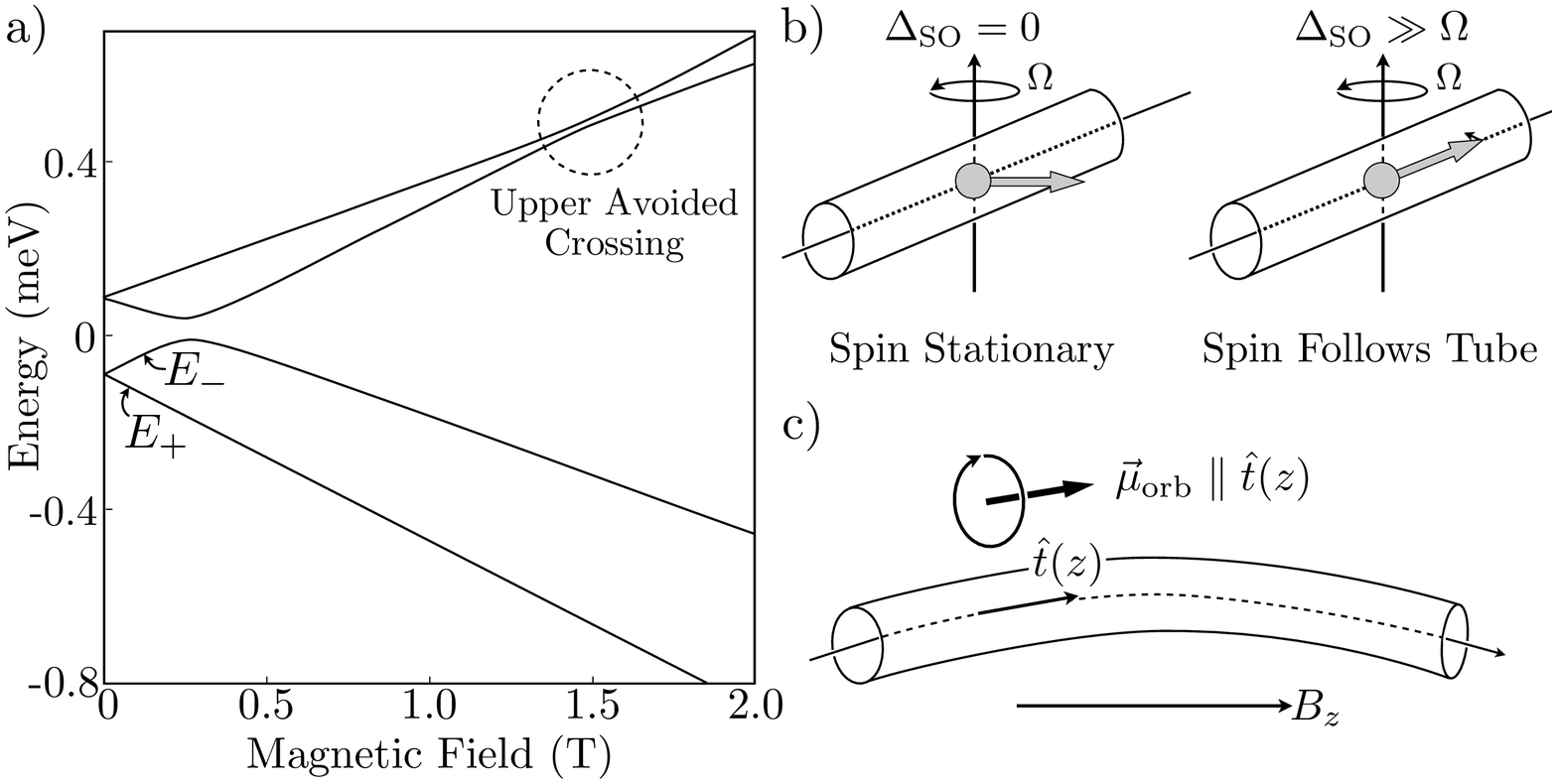}
 \caption[]
{Spin relaxation due to deflection-coupling to bending mode phonons.
a) Single electron quantum dot energy level diagram 
with the parameter values of Ref.~[\onlinecite{Hugh}].
 The four-dimensional subspace is spanned by the states $\Ket{\tau\ s}$ labeled by the valley index $\tau = \pm 1$ and the spin projection $s = \pm 1$ along the laboratory $z$-axis, see Eqs.(\ref{H_d}) and (\ref{H0}).  For the magnetic field $B=0$, the upper (lower) Kramers doublet includes states $\Ket{1,1}$ and $\Ket{-1,-1}$ ($\Ket{1,-1}$ and $\Ket{-1,1})$. 
Due to the flexon density of states singularity as $\omega \rightarrow 0$, we focus on spin relaxation rates between levels with small energy splittings 
in the lower Kramers doublet (denoted by $E_\pm$), and in the narrow upper avoided crossing (dashed circle).
b) Gedanken experiment to illustrate the coupling mechanism.  In the absence of spin-orbit coupling, the electron spin remains fixed in the laboratory frame irrespective of the nanotube's motion.  When the strength of spin-orbit coupling $\Delta_{\rm SO}$ is much greater than the rate of motion $\Omega$, coupling between the spin and orbital moments keeps the electron spin aligned with the direction of the tube axis.
c) For a nanotube with constrained ends, bending mode phonons cause spatial variations in the direction of the nanotube axis $\hat{\bm t}(z)$, which locally couple to the electron spin as described above. 
}
\label{fig1}
\vspace{-4mm}
\end{figure}

As techniques for preparing ultra-clean samples and studying them in cryogenic environments have become available, few-electron quantum dots have emerged as a powerful tool for studying the electron spin properties of nanotubes\cite{Ferdinand,HughNP, Hugh}.
In the experiment by Kuemmeth et al.\cite{Ferdinand}, single electron and single hole quantum dot spectra were shown to display the characteristics of coherent coupling between the electron's spin and its orbital magnetic moment\cite{Ando2000,Minot04,HHGB06,Chico, Jeong09, Izumida,Bulaev}, see Fig.\ref{fig1}a. 
Experiments in the spin-blockade regime of few electron double quantum dots\cite{OnoSB, Buitelaar, Hugh} have also yielded important information about spin relaxation in nanotubes.
In particular, the experiment by Churchill et al.\cite{Hugh} demonstrated a minimum of the spin lifetime $T_1$ near the narrow avoided crossing between levels circled in Fig.\ref{fig1}.
Below we will identify a mechanism of spin relaxation in nanotube quantum dots that is based on the coupling of an electron's spin to the nanotube's deflection.  
This mechanism provides a deeper understanding of the $T_1$ minimum. 
Our theory is developed for a single electron quantum dot, with the corresponding energy level diagram shown in Fig.\ref{fig1}a. 
While we envision that the basic results are more general, a full investigation of the details for many-electron systems is 
beyond the scope of this paper.

In order for spin relaxation to occur, energy must be transferred from the spin to the environment, which for confined electrons typically involves the phonon bath of the host crystal.
Due to strong coupling, phonons play an especially important role in nanotubes\cite{SuzAndo, MarianiOppen, Sapmaz, Leturcq, Cavaliere}.
Crucially for us, clean nanotubes possess quadratically dispersing bending modes\cite{Bulaev, SuzAndo, Mahan, MarianiOppen} with $\omega(q) \propto q^2$.
The corresponding density of states for each such mode has a Van Hove singularity $dq/d\omega \propto 1/\sqrt{\omega}$ at low energies ($\omega\rightarrow 0$).
Thus any relaxation mechanism that relies on coupling to such phonons is expected to become especially efficient for small energy transfers.

Indeed, Bulaev et al.\cite{Bulaev} considered spin-orbit coupling to the bending mode deformation potential, and used the singularity in the density of states to predict a spectacular {\it enhancement} of $1/T_1 \propto 1/\sqrt{\omega}$ near the upper anticrossing point of Fig.\ref{fig1}a in the high temperature regime.
In the low temperature regime, however, this mechanism predicts a {\it suppression} of spin relaxation $1/T_1 \propto \sqrt{\omega}$.
Because the experiments of Ref.[\onlinecite{Hugh}] were performed in an intermediate regime $\hbar\omega/k_B \sim T \sim 100\ {\rm mK}$, it is difficult to confidently link the $T_1$ minimum to the flexon deformation potential coupling.
The deflection-coupling mechanism described below resolves this ambiguity, as it ensures a $T_1$ minimum for all temperatures.

Borysenko et al.\cite{Borysenko} also invoked the density of states singularity, but rather than the deformation potential they considered spin-flexon coupling through the Zeeman interaction due to the g-factor anisotropy $\Delta g$.
This coupling is weak because $\Delta g \approx 10^{-2}$, see Ref.[\onlinecite{Chauvet}] and footnote [\onlinecite{footnoteI}].
Moreover, because the orbital moment\cite{Minot04} $\mu_{\rm orb}$ was not taken into account, the upper avoided crossing did not appear in their spectrum and the theory was only applied to the low-magnetic field Kramers doublets.


Here we consider a different coupling mechanism between flexural modes and electron spin.
Similar to the mechanism of Borysenko et al.\cite{Borysenko}, the coupling is proportional to the nanotube deflection, arising from local changes of the direction of the nanotube axis in the global (laboratory) reference frame.
Unlike their mechanism, however, the spin-orbit interaction of Refs.[\onlinecite{Ferdinand, Hugh}] is intrinsically sufficient to provide the coupling, which thus operates even in 
zero external magnetic field.
Therefore, our mechanism is efficient near both the upper avoided crossing and the narrow Kramers doublets, is not limited by small $\Delta g$, 
and is parametrically stronger than the deformation potential mechanism at low energies (small $q$) because it is proportional to the deflection $\theta(z)$ rather than the deformation $\epsilon = \partial\theta/\partial z\sim q\theta$. 

The physical origin of the coupling mechanism can most easily be understood through a thought experiment, illustrated in Fig.\ref{fig1}b.
Imagine rigidly twirling a nanotube in free space about an axis perpendicular to the tube axis.
In the absence of spin-orbit coupling, an electron spin initially oriented in some direction in the global (laboratory) reference frame will remain polarized in the same direction 
as the nanotube rotates around it.
Now consider the opposite case of strong coupling between the electron spin and orbital magnetic moments.
If the electron is initialized to a state with the orbital and spin moments aligned along the tube axis, then as the tube slowly rotates, the electron spin will adiabatically follow the direction of the tube axis.
For weaker coupling and/or in the presence of an external magnetic field which provides a preferred direction for the electron spin in the laboratory frame, non-trivial spin-dynamics will be generated.


In the more relevant situation of a freely vibrating, electrically-contacted, suspended or substrate-supported nanotube with immobilized ends, bending mode phonons can accomplish the same effect {\it locally}, section-by-section throughout the tube, as indicated in Fig.\ref{fig1}c.
Below we treat this effect perturbatively for small deflections.
To make connection to the Gedanken experiment depicted in Fig.\ref{fig1}c where the spin follows the direction of the tube axis, we note that the electron-flexon coupling Hamiltonian $H_{\rm s-ph}$ [see Eq.(\ref{HSPh}) and accompanying discussion below] used throughout this paper depends only on the instantaneous deflection of the nanotube.
Because the local coupling strength depends on the deflection {\it angle}, it is proportional to the first derivative of the nanotube displacement along the tube axis $z$, and hence to the first power of the phonon momentum $q$.


The plan for the rest of the paper is as follows.
In section \ref{SpinPhononCoupling}, we present the model in detail and define the transition rates to be calculated.
Then in section \ref{LowerKramers} we calculate the spin relaxation rate between states of the field-split lower Kramers doublet for weak external field $B \approx 0$. 
In section \ref{UpperCrossing} we calculate the spin relaxation rate between the two nearly degenerate states at the upper avoided crossing of Fig.\ref{fig1}a and demonstrate the scaling of the $T_1$ minimum in the high-temperature and low-temperature limits. 
The main results of sections \ref{LowerKramers} and \ref{UpperCrossing} are summarized in Eqs.(\ref{WpmFinal}) and (\ref{W12}), respectively.
Last, in section \ref{secDisc} we discuss how this picture is modified in ``dirty'' tubes where a substrate or coating may lead to flexon localization.
 
\section{Spin Phonon Coupling}\label{SpinPhononCoupling}
We consider a single electron confined in a semiconducting (narrow-gap or large-gap) nanotube quantum dot, which in the leading approximation is described by the Hamiltonian
\be
\label{H_d1}H_d = \hbar v_F \left[\tau_3 k_c \sigma_1 - i\sigma_2\frac{d}{dz}\right] + V(z),
\ee
where $v_F \approx 10^6\, {\rm m/s}$ is the Fermi velocity of graphene, $\tau_3$ is the isospin Pauli matrix with eigenvalue $\tau$ in valley $K_\tau$, where $K_1 = 2\pi/3a(1, \sqrt{3})$ and $K_{-1} = -K_1$ are the $K$ and $K'$ points of the graphene Brillouin zone and $a$ is the lattice constant, $\sigma_1$ and $\sigma_2$ are pseudospin (sublattice space) Pauli matrices, and $V(z)$ is a confining potential in the longitudinal direction $z$.
The transverse momentum $\hbar k_c$, which sets the band gap $E_g = 2\hbar v_F k_c$ for free electrons when $V(z) = 0$, is determined by the boundary condition and warping effects that arise from rolling the graphene sheet into a cylinder.
For large-gap nanotubes, $k_c$ is proportional to $1/R$, where $R$ is the radius of the tube.
For narrow-gap tubes, $k_c \propto \cos 3\eta/R^2$ is non-zero due to 
curvature of the graphene sheet\cite{KaneMele}.
Here $\eta$ is the winding angle of the tube\cite{Ando2000, Bulaev, Charlier}.  
To keep the discussion general, for now we avoid imposing any particular form of the longitudinal confining potential.

Due to the 2-fold real spin and 2-fold isospin symmetries, all eigenstates of $H_d$ are four-fold degenerate.
In particular this applies to the ground state: 
\be
\label{H_d}H_d \Ket{\tau\ s} = E_0 \Ket{\tau\ s},
\ee
where $\Ket{\tau\ s} = \Ket{\tau}\otimes\Ket{s}$ is factorized into orbital and spin parts satisfying $\tau_3 \Ket{\tau} = \tau\Ket{\tau}$, $s_z\Ket{s} = s\Ket{s}$.
Here $s_z$ is the real spin Pauli matrix associated with the $z$ direction in the laboratory frame.
Note that $\Ket{\tau}$ describes both the quantized longitudinal and transverse orbital motion of an electron in valley $K_\tau$. 
The specific expression for $E_0$ depends on the structure of the nanotube and on the potential $V(z)$.
Below we set the zero of energy to be $E_0 = 0$.

We now add another piece $H_S$ to the full system Hamiltonian $H = H_d + H_S$, that is central to this paper. 
It includes the coupling of an external magnetic field $\vec{B}$ to the electronic orbital and spin magnetic moments $\mu_{orb}$ and $\mu_B$, spin-orbit coupling, and intervalley scattering,
\be
\label{H_S}H_S = \frac{\Delta_{\rm SO}}{2}\tau_3(\hat{\vec{t}}\cdot\vec{s}) + \Delta_{KK'}\tau_1 - \mu_{orb}\tau_3(\hat{\vec{t}}\cdot\vec{B}) + \mu_B(\vec{s}\cdot\vec{B}),
\ee
where $\hat{\vec{t}} = (t_x, t_y, t_z)$ is the local tangent unit vector at each point along the tube, $\hat{\vec{t}} = \hat{\vec{t}}(z)$, with components defined in the laboratory frame, and $\Delta_{\rm SO}$ and $\Delta_{KK'}$ are the spin-orbit and intervalley coupling matrix elements. 
The constant $\Delta_{\rm SO}$ in our model Hamiltonian $H_S$ absorbs both currently known spin-orbit coupling terms\cite{Ando2000,Chico,Jeong09,Izumida}.
Orbital angular moments were discovered by Minot et al. in Ref.[\onlinecite{Minot04}], and spin-orbit coupling was found and measured by Kuemmeth et al. and Churchill et al. in Refs.[\onlinecite{Ferdinand,Hugh}].
We assume that at rest the tube is straight and oriented along the laboratory $z$ axis.

Coupling to flexural phonons appears when we take into account the fact that the tangent vector is actually an operator depending on the phonon displacement coordinates $\vec{u}(z)$ at each point along the tube.
For small amplitudes, long-wavelength deflections are described by $\hat{\vec{t}}(z) = \hat{\bf z} + d\vec{u}_\perp(z)/dz$, where $\vec{u}_\perp$ is the nanotube displacement perpendicular to $\hat{\bf z}$.
Substituting this expression for $\hat{\vec{t}}$ into Eq.(\ref{H_S}), we split the Hamiltonian $H_S$ into a zero-order unperturbed part\cite{Hugh}
\be
\label{H0} H_0 = \frac{\Delta_{\rm SO}}{2}\tau_3s_z + \Delta_{KK'}\tau_1 - \mu_{orb}\tau_3B_z + \mu_B(\vec{s}\cdot\vec{B}),
\ee
and a spin-phonon coupling part
\be
\label{HSPh} H_{\rm s-ph} = \frac{\Delta_{\rm SO}}{2}\left(\frac{d\vec{u}_\perp}{dz}\cdot\vec{s}\right)\tau_3 - \mu_{orb}\left(\frac{d\vec{u}_\perp}{dz}\cdot\vec{B}\right)\tau_3.
\ee
As usual in the theory of condensed matter systems, and in the spirit of the Born-Oppenheimer approximation, the spin-phonon coupling Hamiltonian $H_{\rm s-ph}$ depends only on the instantaneous deflection $\vec{u}_\perp(z)$.
In terms of the flexon creation and annihilation operators $a^\dagger_{q\alpha}$ and $a_{q\alpha}$,
\be
\frac{d\vec{u}_\perp}{dz}(z)=\sum_{q\alpha}iq\sqrt{\frac{{\hbar}}{2\rho L_z\omega_q}}{\hat{\bf x}}_\alpha(a_{q\alpha}+a^\dagger_{q\alpha})e^{iqz},
\label{PhononDisplacement}
\ee
where $\alpha$ distinguishes two bending mode polarizations $\hat{{\bf x}}_\alpha$, $L_z$ is the nanotube 
length, 
$q$ and $\omega_q$ are flexon momentum and frequency, and $\rho$ is mass per unit length. 
As shown in Refs.[\onlinecite{MarianiOppen,Mahan,SuzAndo}], $\omega(q)=\beta q^2$ for small $q$; because of the nanotube symmetry, the two polarization modes are degenerate.

The phonon normal mode profiles and frequencies depend in a non-universal way on the details of the nanotube's environment (e.g. on 
boundary conditions, and on 
the presence of disorder or coating particles on the tube's surface).
In Eq.(\ref{PhononDisplacement}) we capture the generic behavior of the mechanism by assuming plane-wave normal modes. 
However, below we write the main results (\ref{WpmFinal}) and (\ref{W12}) in terms of a form-factor $M(q)$, Eq.~(\ref{Mq}), which can be generalized to other normal mode profiles. 
Indeed, a more general discussion of the flexon normal modes is provided in Sec.\ref{secDisc}. 

Similar comments can be made regarding the influence of electron-electron interactions in many-electron quantum dots. 
A homogeneous background of closed electron shells should merely renormalize the coefficients without influencing the basic qualitative results. 
However, a strongly inhomogeneous electron background including charge puddles may influence electron-flexon coupling more profoundly. 
Such a regime is outside the scope of our investigation.

The spin Hamiltonian $H_S$ is defined on the four-dimensional subspace spanned by the eigenstates $\Ket{\tau\ s}$ of $H_d$, Eq.(\ref{H_d}).
The orbital term $\mu_{orb}\tau_3 (\hat{\vec{t}}\cdot\vec{B}) $ results from a shift of $k_c$ due to the vector potential associated with $\vec{B}$.
This renormalizes the gap $E_g$, which in principle affects the longitudinal confinement (see Appendix \ref{AppOverlap}).
Because this shift is small in the regime of interest for experiments [\onlinecite{Ferdinand,Hugh}], $\mu_{orb} (\hat{\vec{t}}\cdot\vec{B})/E_g \lesssim 1/30$, we ignore its effect on the longitudinal motion. 
By ignoring coupling to higher orbital levels we omit terms which are small in the inverse level spacing, 
but in return obtain general analytical formulas which allow us to clearly extract the essential physics underlying the deflection coupling mechanism.

Armed with the perturbation (\ref{HSPh}), we calculate transition rates between eigenstates of the zero-order Hamiltonian $H_d + H_0$ using Fermi's Golden Rule:
\be
\label{W} W_{fi} =\frac{2\pi}{\hbar^2} \overline{|\MatEl{\psi_f}{H_{\rm s-ph}}{\psi_i}|^2}\frac{L_z}{2\pi}\left.\frac{dq}{d\omega}\right\vert_{\hbar\omega = E_f - E_i}.
\ee
Here $\Ket{\psi_i}$ and $\Ket{\psi_f}$ are the initial and final states satisfying $[H_d + H_0]\Ket{\psi_n} = E_n\Ket{\psi_n}$, and the over-bar indicates averaging over the thermal phonon distribution.

As discussed further in Appendix \ref{AppOverlap}, due to the fact that the orbital states $\Ket{\tau}$ with $\tau = \pm 1$ are time-reversal conjugate, 
the density $n(z_0) \equiv \MatEl{\tau}{\delta(z - z_0)}{\tau}$, which 
is a {\it real} scalar, is independent of the isospin index $\tau$.
Using this fact and the property that Eq.(\ref{HSPh}) is diagonal in the isospin index $\tau$, 
any matrix element $\MatEl{\psi_f}{H_{\rm s-ph}}{\psi_i}$ includes the longitudinal form factor 
\be
\label{Mq}M(q) \equiv \int_{-\infty}^{\infty}dz\,n(z)\, e^{iqz},
\ee
which depends on the specific form of the longitudinal confinement, but otherwise does not depend on the composition of $\Ket{\psi_i}$ and $\Ket{\psi_f}$ in terms of the basis states $\{\Ket{\tau\ s}\}$.

In Appendix \ref{AppOverlap} we calculate $M(q)$ for square-well confinement.
For small momentum $q \rightarrow 0$, the form factor $M(q)\rightarrow 1$, simply reflecting the wave function normalization.
For large phonon momentum $qL_d \gg 2\pi$, where $L_d$ is the electronic length of the quantum dot, the factor $e^{iqz}$ oscillates rapidly relative to the electron wave function and leads to a strong suppression of the matrix element, as will be discussed in section \ref{secDisc}.  
Note that we distinguish between $L_z$, the unconstrained length of the nanotube relevant for phonon normal modes, and $L_d$, the quantization length of the electronic wavefunction controlled by electrostatic gates, which in general are independent quantities (see e.g. Ref.[\onlinecite{Cavaliere}]).

\section{Relaxation rate in the lower Kramers doublet}\label{LowerKramers}
Due to the flexon density of states singularity as $\omega \rightarrow 0$, we focus our attention on spin relaxation between states with small energy splittings.
We begin by considering relaxation between the lower Kramers pair of states (see Fig.\ref{fig1}a) in a small applied magnetic field.
Although time reversal symmetry prevents relaxation from occurring at $B \equiv |\vec{B}| = 0$ (Van Vleck cancellation, see Refs.[\onlinecite{VanVleck, RashbaSheka, KN, nonadiab}]), the large density of states contributes to a steep rise of the relaxation rate as $B$ is increased away from zero.

To apply Eq.(\ref{W}) for the relaxation rate, the first step is to find the two lowest energy eigenstates of Hamiltonian $H_0$. 
We take $\Delta_{\rm SO} > 0$, as observed in Ref.[\onlinecite{Ferdinand}].
Diagonalization is accomplished in two steps: we first apply a unitary transformation
\begin{widetext} 
\bea
\label{U} U=\frac12\sqrt{{\frac{\Delta+\Delta_{\rm SO}}{2\Delta}}}~
\left[\left(1-\frac{\Delta_{KK'}}{|\Delta_{KK'}|}\right)+
s_z\tau_3\left(1+\frac{\Delta_{KK'}}{|\Delta_{KK'}|}\right)\right]
+{\frac12}\sqrt{\frac{{\Delta-\Delta_{\rm SO}}}{2\Delta}}\left[\left(1+\frac{\Delta_{KK'}}{|\Delta_{KK'}|}\right)
-is_z\tau_2\left(1-\frac{\Delta_{KK'}}{|\Delta_{KK'}|}\right)\right]
\eea
\end{widetext}
to switch to a frame where $UH_0(B = 0)U^\dagger = -\frac{\Delta}{2}s_z\tau_3$ is diagonal.
Here $\Delta \equiv \sqrt{\Delta_{\rm SO}^2 + 4 \Delta_{KK'}^2}$.
For $\Delta_{\rm SO} < 0$, a different unitary transformation must be chosen.

After the transformation, the zero-order Hamiltonian $H_0^U = UH_0U^\dagger$ for $B \neq 0$ reads
\begin{widetext}
\be
\label{H0U}H_0^U =\!\left(-\frac{\Delta}{2}\tau_3s_z-\mu_{orb}B_z\frac{\Delta_{\rm SO}}{\Delta}\tau_3+\mu_BB_zs_z\right)-2\mu_BB_\perp\frac{\Delta_{KK'}}{\Delta}\tau_2s_y-\mu_BB_\perp\frac{\Delta_{KK'}\Delta_{\rm SO}}{|\Delta_{KK'}|\Delta}s_x
-2\mu_{orb}B_z\frac{|\Delta_{KK'}|}{\Delta}\tau_1s_z,
\ee
\end{widetext}
where without loss of generality we choose the perpendicular component of the magnetic field $\vec{B}_\perp$ to be oriented along the $\hat{\bf x}$ direction, $B_x = B_\perp, B_y = 0$.

To linear order in $B/\Delta$, the lowest two energy eigenvalues are $E_\pm = -(\frac{\Delta}{2} \pm \Lambda)$, with 
\be
\label{Epm}  \Lambda = \frac{1}{\Delta}\sqrt{(\mu_{orb}\Delta_{\rm SO}+\mu_B\Delta)^2B_z^2+4\mu_B^2B_\perp^2\Delta_{KK'}^2}.
\ee
Explicit expressions for the corresponding four-spinor eigenvectors $\Ket{\psi_\pm}$ are too cumbersome to reproduce here, but will be used below to calculate the necessary matrix elements. 
In the case $\Delta_{KK'} \rightarrow 0$, $\Ket{\psi_+}$ and $\Ket{\psi_-}$ correspond to states adiabatically connected to $\Ket{+ \downarrow}$ and $\Ket{- \uparrow}$ as $B \rightarrow 0$, where $+(-)$ indicates $\tau = 1$ ($\tau = -1$).
More generally, the $\pm$ label in $\Ket{\psi_\pm}$ simply indexes the two states which are degenerate as $B\rightarrow 0$.

Substituting the expressions for $\Ket{\psi_\pm}$ into Eq.(\ref{W}) yields the 
rate $W^E_{\alpha}$ ($W^A_{\alpha}$) 
for emission (absorption) of a phonon with momentum $q$ and polarization $\alpha$
\bea
\label{Wpm} W^E_{\alpha} &=&\frac{\pi q^2}{\hbar\rho L_z\omega_q}(N_q + 1)|\MatEl{\psi_+}{H^U_\alpha}{\psi_-}|^2\frac{L_z}{2\pi}\frac{dq}{d\omega},\nonumber\\
 W^A_{\alpha} &=&\frac{\pi q^2}{\hbar\rho L_z\omega_q}N_q|\MatEl{\psi_-}{H^U_\alpha}{\psi_+}|^2\frac{L_z}{2\pi}\frac{dq}{d\omega},
\eea
where $N_q$ is the thermal occupation number for the phonon with $\hbar\omega_q = E_- - E_+$, and 
\be
H^U_\alpha = Ue^{iqz}\!\!\left[\frac{\Delta_{\rm SO}}{2}\left(\hat{\bf x}_\alpha\cdot\vec{s}\right)\tau_3 - \mu_{orb}\left(\hat{\bf x}_\alpha\cdot\vec{B}\right)\tau_3\right]U^\dagger
\ee
is the $U$-transformed spin-phonon coupling operator of Eqs.(\ref{HSPh}) and (\ref{PhononDisplacement}) with the phonon operators and related dimensional constants removed. 
Using the explicit expressions for $\Ket{\psi_\pm}$, we find
\bea
\MatEl{\psi_+\!}{H^U_x}{\psi_-\!}\!\! 
&=&- \frac{2|\Delta_{KK'}|}{\Delta^3\Lambda}\,\bigg[(\mu_{orb}\Delta^2-\mu_B\Delta_{\rm SO}^2)\mu_BB_\perp^2 \nonumber\\
&&+\ \Delta_{\rm SO}(\mu_{orb}\Delta_{\rm SO}+\mu_B\Delta)\mu_{orb}B_z^2\bigg]M(q)\nonumber\\
\MatEl{\psi_+\!}{H^U_y}{\psi_-\!}\!\! &=& \frac{2i|\Delta_{KK'}|}{\Delta^2}M(q)\Delta_{\rm SO}\mu_{orb}B_z,
\label{KMats}
\eea
where $M(q)$ is the longitudinal form factor calculated in Appendix \ref{AppOverlap}.

First, because $\Lambda \propto B$, it is apparent that both matrix elements in (\ref{KMats}) vanish as $B\rightarrow 0$ as expected for a transition between time-conjugate partners of a Kramers doublet.
Second, both matrix elements vanish when $\Delta_{KK'} \rightarrow 0$, which is also expected because in this limit the two members of the Kramers doublet belong to opposite valleys $K$ and $K'$, which are not coupled by long wavelength phonons.
Finally, although it is not immediately obvious from Eq.(\ref{KMats}), in the limit $B_\perp \rightarrow 0$ the matrix elements for transitions involving the two polarization modes are equal in magnitude as required by axial symmetry: $|\MatEl{\psi_+}{H^U_x}{\psi_-}| = |\MatEl{\psi_+}{H^U_y}{\psi_-}| = 2|M(q)\Delta_{\rm SO}\Delta_{KK'}\mu_{orb}B_z|/\Delta^2$.

The relaxation rate $\Gamma = 1/T_{1}$ 
is the {\it total} rate of phonon emission and absorption in both channels $\hat{\bf x}_\alpha$:
\be
\label{EqT1} \Gamma = \sum_\alpha \left(W_{\alpha}^E + W_{\alpha}^A\right).
\ee
To simplify the complicated expressions of Eq.(\ref{KMats}), we take advantage of the fact that
\be
\label{SmallMuB} |\mu_B/\mu_{orb}| \ll 1,
\ee 
typically\cite{HughNP,Ferdinand,Hugh} $\mu_B \approx 0.1\mu_{\rm orb}$, and that $B_\perp$ always appears multiplied by $\mu_B$, and so retain only the the leading term in $\mu_{orb}B_z$. 
Using the energy conservation law $\hbar\omega_Z \equiv E_- - E_+ = \hbar\omega_q \approx 2\mu_{orb}B_z$ and $d\omega/dq = 2\beta q$, 
expression (\ref{Wpm}) evaluates to 
\be
\frac{1}{T_{1}} \approx 
  \frac{\hbar |M(q)|^2}{2\rho}\!\!\left(\frac{\Delta_{\rm SO}\Delta_{KK'}}{\Delta^2}\right)^2\!\left(\frac{\omega_Z}{\beta}\right)^{3/2}\!\!\!\coth \left(\frac{\hbar\omega_Z}{2k_BT}\right).
\label{WpmFinal}
\ee

Not surprisingly, the $\omega_Z$ dependence of expression (\ref{WpmFinal}) has the same form as that found by Borysenko et al.\cite{Borysenko} 
However, their relaxation rate $\Gamma_{\Delta g}$ is suppressed by a very weak g-factor anisotropy $\Delta g/g \approx 0.01$\cite{Chauvet}.
Comparing prefactors, we find that the rates of relaxation due to these two mechanisms are related by the ratio
\be
\frac{\Gamma_{\Delta g}}{\Gamma} = \frac12\left(\frac{\Delta g}{g}\cdot\frac{\Delta^2}{\Delta_{\rm SO}\Delta_{KK'}}\right)^2.
\ee
For parameters in the regime which appears to be relevant for experiments [\onlinecite{Ferdinand,Hugh}], i.e. $\Delta_{\rm SO} \gtrsim \Delta_{KK'}$, we find $\Gamma_{\Delta g}/\Gamma \ll 1$, indicating that the spin-orbit mediated mechanism studied herein should strongly dominate the behavior at low fields.

We now provide an estimate of $\Gamma$ based on Eq.(\ref{WpmFinal}), with realistic experimental parameters.
For $\beta$, the coefficient of $q^2$ in the flexon dispersion, we use the result of Mariani and von Oppen\cite{MarianiOppen} 
\be
\beta = R\sqrt{\frac{2\mu(\mu + \lambda)}{(2\mu + \lambda)\rho_s}},
\ee
where $R \approx 1\ {\rm nm}$ is the nanotube radius\cite{Hugh}, $\rho_s = 10^{-7}$g/cm$^2$ is the graphene sheet density\cite{Borysenko}, related to the linear density $\rho$ through  $\rho = 2\pi R \rho_s$, and $\mu \approx 4\lambda = 9\ {\rm eV \AA}^{-2}$ are Lam\'{e} coefficients\cite{MarianiOppen}.
This gives $\hbar\beta \approx 10^4\ \mu{\rm eV\, nm}^2$. 
Note that the theory of flexon vibrations in nanotubes was first studied independently by Suzuura and Ando\cite{SuzAndo} and Mahan\cite{Mahan}, who used different models of the curved graphene sheet.
Mahan's theory and that of Suzuura and Ando, which was later applied by Bulaev et al.\cite{Bulaev} and by Mariani and von Oppen\cite{MarianiOppen}, predict different functional dependence of $\beta$ on $\lambda$ and $\mu$.
However, the numerical values obtained from each of these theories 
differ only by an insignificant factor of order 1.

In the low temperature limit, $\Gamma$ is suppressed as $\omega_Z^{3/2}$ as $B\rightarrow 0$. 
The ratio of $\Delta_{\rm SO}$ to $\Delta_{KK'}$ is strongly sample-dependent; for illustration we take $\Delta_{\rm SO}/\Delta_{KK'} = 2$. 
For a typical field $B = 10^3\ {\rm G}$ nominally oriented along the tube axis, we find $\Gamma \approx 300\ {\rm s}^{-1}$ for $T = 0$.
For high temperatures the relaxation rate is enhanced by $k_BT/\hbar\omega_Z$ due to the presence of thermal phonons, and the $\omega_Z\rightarrow 0$ limit of Eq.(\ref{WpmFinal}) goes as $\Gamma \propto \omega_Z^{1/2}$.

In this section we focused on relaxation within the lowest Kramers doublet.
When condition (\ref{SmallMuB}) is satisfied, however, the results above can also be applied directly to relaxation within the upper Kramers doublet.

\section{Upper avoided crossing}\label{UpperCrossing}

We now turn our attention to spin relaxation in the vicinity of the upper avoided crossing circled in Fig.\ref{fig1}a, where a minimum in $T_1(B)$ was discovered by Churchill et al.\cite{Hugh}. 
Such an acceleration of spin-relaxation in the high temperature regime was first predicted in Ref.[\onlinecite{Bulaev}] for the case of electron-phonon coupling through the deformation potential.
Here we build on the idea outlined in that work, and consider how relaxation is affected in the presence of the deflection-coupling mechanism described above. 
As we will show below, for small level splittings $\omega$ the rate is parametrically enhanced by a factor of $1/\omega$, thus producing a robust enhancement of spin relaxation both in the high and low temperature regimes.

Because the secular equation corresponding to Hamiltonian (\ref{H0}) is fourth order, the spectrum of the system for moderate values of $B$ cannot in general be described in any simple form.
Furthermore, depending on whether or not the system possesses certain symmetries such as valley conservation ($\Delta_{KK'} = 0$) or axial symmetry ($B_\perp = 0$), level crossings may be exact or avoided.
The various possible regimes are reviewed in subsection \ref{secSpectrum} below (see also Ref.[\onlinecite{Bulaev}]).
In subsection \ref{secUCRates} we calculate the relaxation rate close to the upper avoided crossing for the general case $\Delta_{KK'} \neq 0, B_\perp \neq 0$, and then provide results for the special limiting cases $\Delta_{KK'} = 0$ and $B_\perp = 0$ in subsection \ref{secUCLimits}.

\subsection{Spectrum for $B \neq 0$}\label{secSpectrum}

For $B > 0$, the spectrum of Hamiltonian (\ref{H0}) displays two avoided level crossings, shown in Fig.\ref{fig1}a.
Coupling between nominally degenerate states is provided by intervalley scattering $\Delta_{KK'}$ or by the magnetic field component perpendicular to the tube axis, $B_\perp$. 
To provide a better understanding of what controls the splitting at each crossing individually, we first examine the two limiting cases $\Delta_{KK'} = 0, B_\perp \neq 0$ and $\Delta_{KK'}\neq 0, B_\perp = 0$, which are described by simple analytical expressions. 

\begin{figure}
\includegraphics[width=3.2in]{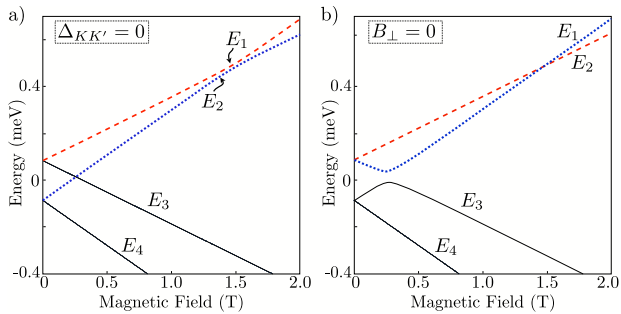}
 \caption[]
{(color online) Behavior of the spectrum in limiting regimes $\Delta_{KK'} = 0$ or $B_\perp = 0$.
a) Spectrum for $\Delta_{KK'} = 0, B_\perp \neq 0$.  The levels $E_1$ and $E_2$ exhibit an avoided crossing with splitting controlled by the field misalignment $B_\perp$.  The crossing between $E_2$ and $E_3$ is exact, except when the field direction deviates very far from the tube axis (see text).
b) Spectrum for $\Delta_{KK'} \neq 0, B_\perp = 0$.  Intervalley scattering opens an avoided crossing between levels $E_1$ and $E_3$.  The exact crossing between $E_1$ and $E_2$ persists as long as $\Delta_{\rm SO} \gtrsim 2(\mu_B/\mu_{orb})\Delta_{KK'}$.
}
\label{FigUCSpectra}
\vspace{-4mm}
\end{figure}
In the first case, $\Delta_{KK'} = 0, B_\perp \neq 0$, the four branches of the spectrum are given by
\bea
\label{DKK0}
E_1 =\!&\!\! \mu_{orb}B_z\!\!\!\!\! &\!+ \frac12 \sqrt{(\Delta_{\rm SO} - 2\mu_BB_z)^2 + 4\mu_B^2B_\perp^2}\nonumber\\
E_2 =\!&\!\! \mu_{orb}B_z\!\!\!\!\!  &\!- \frac12 \sqrt{(\Delta_{\rm SO} - 2\mu_BB_z)^2 + 4\mu_B^2B_\perp^2}\nonumber\\
E_3 =\!&\!\! -\mu_{orb}B_z\! &\!+ \frac12 \sqrt{(\Delta_{\rm SO} + 2\mu_BB_z)^2 + 4\mu_B^2B_\perp^2}\nonumber\\
E_4 =\!&\!\! -\mu_{orb}B_z\! &\!- \frac12 \sqrt{(\Delta_{\rm SO} + 2\mu_BB_z)^2 + 4\mu_B^2B_\perp^2},
\eea
shown in Fig.\ref{FigUCSpectra}a.
An exact crossing at the upper intersection can only occur if $E_1 = E_2$, which implies $2\mu_B B_Z = \Delta_{\rm SO}, B_\perp = 0$; therefore the spectrum displays an avoided crossing for any $B_\perp \neq 0$.
Similarly, an exact crossing at the lower (central) intersection occurs when $E_2 = E_3$, which requires
\be
\label{MiddleCrossingCondition} 4\mu^2_{orb}B^2\left[\cos^2\phi - \frac{\mu^2_B}{\mu^2_{orb} - \mu^2_B}\sin^2\phi\right] = \Delta_{\rm SO}^2,
\ee
where $\phi$ is the angle between the magnetic field $\vec{B}$ and the tube axis $\hat{\bf z}$.
This condition is always satisfied for some value of $B$ when the magnetic field direction does not deviate too far from the tube axis, $\tan^2\phi < (\mu_{orb}/\mu_B)^2 - 1 \approx 100$.

In the opposite case, $\Delta_{KK'} \neq 0, B_\perp = 0$, we find
\bea
\label{Bp0}
E_1 =\!&\!\! \mu_B B_z\!\!\!\!\! &\! + \frac12\sqrt{(\Delta_{\rm SO} - 2\mu_{orb}B_z)^2 + 4\Delta_{KK'}^2}\nonumber\\
E_2 =\!&\!\! -\mu_B B_z\!\! &\! + \frac12\sqrt{(\Delta_{\rm SO} + 2\mu_{orb}B_z)^2 + 4\Delta_{KK'}^2}\nonumber\\
E_3 =\!&\!\! \mu_B B_z\!\!\!\!\! &\! - \frac12\sqrt{(\Delta_{\rm SO} - 2\mu_{orb}B_z)^2 + 4\Delta_{KK'}^2}\nonumber\\
E_4 =\!&\!\! -\mu_B B_z\!\! &\! - \frac12\sqrt{(\Delta_{\rm SO} + 2\mu_{orb}B_z)^2 + 4\Delta_{KK'}^2}.
\eea
An exact crossing $E_1 = E_2$ exists at $B_z$ satisfying
\be
\label{TrueCross} 4\mu_B^2\left[B_z^2 + \frac{\Delta^2_{KK'}}{\mu^2_{orb} - \mu^2_B}\right] = \Delta^2_{\rm SO},
\ee
whenever $4\mu^2_B\Delta^2_{KK'} < (\mu^2_{orb} - \mu_B^2)\Delta^2_{\rm SO}$.
Because $\mu_B \ll \mu_{orb}$, this condition is unlikely to be violated in clean samples.
The gap between $E_1$ and $E_3$ persists for any $\Delta_{KK'} \neq 0$.

More generally, both $\Delta_{KK'},B_\perp \neq 0$, as in Fig.\ref{fig1}a.
Here we gain further insight by focusing on the regime 
\be
\label{CrossCond} 2\mu_BB_z = \Delta_{\rm SO} - \delta\epsilon, 
\ee
where $\delta\epsilon$ is a small deviation from the upper crossing found in the $\Delta_{KK'},B_\perp = 0$ limit.
For small $\delta\epsilon$, $\Delta_{KK'}$, and $B_\perp$, we diagonalize the Hamiltonian within the $2\times 2$ subspace defined by the crossing levels, taking into account coupling to the lower two levels to first order in perturbation theory.
The perturbative treatment is justified by the large energy denominators compared to the coupling matrix elements, $(\mu_{orb} \pm \mu_B)B_z \gg \delta\epsilon, \mu_BB_\perp, \Delta_{KK'}$.
Because $\Delta_{KK'}$ shifts the level crossing point, which is otherwise determined by condition (\ref{CrossCond}), we also require $\Delta_{KK'} \ll (\mu_{orb}/\mu_B)\Delta_{\rm SO}$ to ensure that the perturbation criteria are not violated.
This latter restriction is rather mild, however, due to $\mu_{orb} \gg \mu_B$.

This treatment yields eigenvectors $\Ket{\psi_1}$ and $\Ket{\psi_2}$ (not shown) and eigenvalues
\be
\label{E12} E_{1,2} = \mu_{orb}B_z + \frac{\mu_{orb}\Delta_{KK'}^2 \pm \mathcal{R}}{2(\mu_{orb}^2 - \mu_B^2)B_z},
\ee
where $+(-)$ is used for $E_{1(2)}$, see Fig.\ref{FigUCSpectra}a.
Here
\bea
\label{E12R}\mathcal{R}^2 &=& \left[(\mu_{orb}^2 - \mu_B^2)B_z\delta\epsilon - \mu_B\Delta^2_{KK'}\right]^2\nonumber\\ 
&&+\ 4(\mu^2_{orb} - \mu_B^2)^2B^2_z\mu^2_BB^2_\perp.
\eea
In writing $\mathcal{R}$, we use Eq.(\ref{CrossCond}) and keep only the lowest order term in $\delta\epsilon$.
In the limits $\Delta_{KK'} = 0$ and/or $B_\perp = 0$, Eqs.(\ref{E12}) and (\ref{E12R}) reduce to the earlier results (\ref{DKK0}) and (\ref{Bp0}).

\subsection{Transition Rates for $\Delta_{KK'},B_\perp \neq 0$}\label{secUCRates}
Proceeding as in section \ref{LowerKramers}, the amplitudes for transitions from $\Ket{\psi_1}$ to $\Ket{\psi_2}$ due to coupling to phonons with polarization $\alpha$ are
\bea
\MatEl{\psi_2\!}{H^U_x}{\psi_1\!}\!\! 
&=&\!\! \frac{\Delta_{\rm SO}}{2\mathcal{R}}M(q)\!\left[\mu_B\Delta^2_{KK'} - (\mu^2_{orb} - \mu^2_B)B_z\delta\epsilon\right],\nonumber\\
\MatEl{\psi_2\!}{H^U_y}{\psi_1\!}\!\! &=&\!\! -i\frac{\Delta_{\rm SO}}{2}M(q),
\label{UCMats}
\eea
with $\mathcal{R}$ as defined in Eq.(\ref{E12R}).
In the expressions above, small corrections proportional to $(B_\perp/B_z)^2$ and $\Delta_{KK'}^2/[4(\mu^2_{orb} - \mu_B^2)B_z^2]$ have been omitted.

Interestingly, the contribution from $y$-polarized phonons is constant near the avoided crossing, with an amplitude equal to $M(q)\Delta_{\rm SO}/2$.
When $B_\perp = 0$, the two amplitudes are equal as required by axial symmetry; although the energy splitting $E_1 - E_2$ goes through zero in this limit, the amplitude remains finite.
Unlike the case of the Kramers doublets here there is no time reversal symmetry to prevent such nonzero amplitude.

Additionally, for $B_\perp \neq 0$, the amplitude for emitting an $x$-polarized phonon vanishes for the detuning $\delta\epsilon = \delta\epsilon_0$ satisfying
\be
\delta\epsilon_0 = \frac{\mu_B\Delta^2_{KK'}}{(\mu^2_{orb} - \mu^2_B)B_z}.
\ee
Thus the coupling (and hence the relaxation rate) is asymmetric in detuning relative to the position of the anticrossing.
For $\Delta_{KK'}/\Delta_{\rm SO} \ll 1$, the asymmetry is rather small.
If it can be detected, however, it will allow the value of $\Delta_{KK'}$ to be extracted.

It is convenient to express $\mathcal{R}$ in terms of $\delta\epsilon_0$,
\bea
\mathcal{R} &=& (\mu_{orb}^2 - \mu_B^2)B_z\,\epsilon(B_z,B_\perp),\nonumber \\
\epsilon(B_z,B_\perp) &=& \sqrt{[\delta\epsilon(B_z) - \delta\epsilon_0]^2 + 4\mu_B^2B_\perp^2}.
\eea
Here the energy splitting $E_1 - E_2 = \mathcal{R}/[(\mu^2_{orb} - \mu^2_B)B_z] = \epsilon(B_z,B_\perp)$ is controlled by a combination of $B_z$ (detuning) and $B_\perp$ (avoided crossing gap). 
Summing the squares of the matrix elements (\ref{UCMats}), using the energy conservation law $\hbar\omega_q =  \epsilon(B_z, B_\perp)$, and using the Golden Rule, Eq.(\ref{W}), 
\be
 \label{W12} \Gamma(\epsilon) = \frac{\Delta^2_{\rm SO}|M(q)|^2}{16\, \beta^{3/2}\sqrt{\hbar\epsilon}\rho}\left[1 + \left(\frac{\delta\epsilon - \delta\epsilon_0}{\epsilon}\right)^2\right]\coth\left(\frac{\epsilon}{2k_BT}\right),
\ee
where $\Gamma = \sum_\alpha (W^E_{\alpha} + W^A_{\alpha}) = 1/T_1$ is the total rate of transitions between $\Ket{\psi_1}$ and $\Ket{\psi_2}$.

As expected, the relaxation rate (\ref{W12}) between states near the upper avoided crossing shows a resonant enhancement proportional to $\epsilon^{-1/2}$ ($\epsilon^{-3/2}$) in the low (high) temperature regime due to the combination of the divergent flexon density of states at small energies and the deflection coupling mechanism.
The singularity in the rate is a factor of $1/\epsilon$ stronger than that found by Bulaev et al.\cite{Bulaev} for the deformation potential coupling mechanism.
For larger values of $\epsilon$, the competition between the two mechanisms is sensitive to various system parameters.
Without an analytical expression for $\Gamma(\epsilon)$ for the deformation coupling mechanism, however, it is difficult to make a more detailed comparison in this regime. 

Using the parameters of Ref.[\onlinecite{Hugh}], $\Delta_{\rm SO} = 170\ \mu{\rm eV}$, $\epsilon \approx 10\ \mu{\rm eV}$, and $T \approx 0.1\ K$, we find $\Gamma \approx 10^4 s^{-1}$.
Despite the strong singularity for small energy splittings, this value is still 2 orders of magnitude smaller than the observed relaxation rate.
However, the rate is highly sensitive to parameter values, including the flexon temperature and especially the avoided crossing splitting $\epsilon$. 
The latter depends on the angle of misalignment between the magnetic field and the tube axis, which was estimated to be 5$^\circ$ by electron micrograph, but if smaller could easily lead to enhancement of the rate.
Thus although the magnetic field dependence of $T_1$ from this model supports the association of the $T_1$ minimum observed by Churchill et al.\cite{Hugh} with coupling to flexons, additional experiments are needed to understand the quantitative details of this relationship.
%

\subsection{Limiting cases $\Delta_{KK'} = 0, B_\perp \neq 0$ and $\Delta_{KK'}\neq 0, B_\perp = 0$}\label{secUCLimits}
To gain a fuller picture of the behavior near the upper avoided crossing, we now examine the simple limits $\Delta_{KK'} = 0$ and $B_\perp = 0$ of section \ref{secSpectrum}.
In the limit $\Delta_{KK'} = 0$, $B_\perp \neq 0$,  $E_1$ and $E_2$ exhibit an avoided crossing of splitting $E_1 - E_2 = 2\mu_BB_\perp$. 
At this point, $\delta\epsilon = \delta\epsilon_0$ and therefore $\MatEl{\psi_2}{H^U_x}{\psi_1}$ vanishes; only flexons with $y$-polarization contribute to relaxation.
The finite energy splitting $\epsilon$ at the avoided crossing, controlled by $B_\perp$,  provides a low-energy cutoff for the denominator in Eq.(\ref{W12}).
Further away from the avoided crossing, the rate is suppressed by the decaying phonon density of states, but gains an extra factor of 2 from the phonons with $x$-polarization that become active there.

In the other limit, when $\Delta_{KK'} \neq 0$ and $B_\perp = 0$, there is an exact crossing when condition (\ref{TrueCross}) is satisfied.
Here both flexon polarizations contribute equal transition matrix elements $\Delta_{\rm SO}/2$.
With constant matrix elements in the numerator and no low energy cutoff in the denominator, expression (\ref{W12}) diverges as $B$ approaches the critical value defined by Eq.(\ref{TrueCross}).
Of course, in any real experiment, the divergence will always be cut off by something, such as a small misalignment of the field axis.
However, as found for the general case in the previous subsection, the behavior in all regimes displays an acceleration of spin-relaxation near the upper avoided crossing.
This is a general property of the deflection coupling mechanism of spin-relaxation.

\section{Discussion and Conclusions}\label{secDisc}
Above, we found that the direct coupling between an electron's spin and the deflection of its host nanotube provides an efficient mechanism for electron spin relaxation near narrow avoided level crossings where the energy transfer is small.
Efficient relaxation is made possible by the diverging density of states for bending mode phonons with quadratic dispersion\cite{SuzAndo,Mahan,Bulaev,MarianiOppen}, $\omega_q \propto q^2$.
Such a dispersion law is predicted for clean systems such as perfect suspended nanotubes.
What happens for dirty systems where the nanotube is placed on a substrate, and/or covered by an irregular coating left over from sample processing?

The primary effects of disorder are twofold.
Constraints on the tube's motion alter the displacement profiles of the normal modes.
As a result, the spectrum of normal mode frequencies is altered, with spectral density at very low frequencies generally getting redistributed to higher frequencies.
We begin this section by discussing normal mode profiles and providing some additional discussion of the longitudinal form factor $M(q)$ that appears throughout the text.
Then we discuss how changes in $M(q)$ and $dq/d\omega$ due to flexon localization may affect the relaxation rates calculated above. 

By writing Eq.(\ref{PhononDisplacement}) for the nanotube deflection, we assume that the flexon normal modes are described by plane waves $e^{iqz}$.
Even for a perfectly clean system, however, the boundary conditions at the mechanically constrained points of the tube can mix modes with momenta $\pm q$ to yield normal mode profiles which are sums of sines and cosines.
While these boundary conditions determine the nanotube length $L_z$, the electronic length $L_d$ is determined by a different set of constraints arising from electrostatic potentials created by external gates and/or impurities; thus it is not possible to find a completely general result for the form factor $M(q)$, which depends on the shapes and relative displacements of both the electron wavefunction and
flexon normal mode profile.
Indeed the authors of Ref.[\onlinecite{Cavaliere}] concluded that asymmetric suppression of vibrational sidebands in transport through a nanotube quantum dot could be explained by the presence of a vibrational mode localized near one end of the quantum dot.
Thus we see that the general situation can be quite complicated; we consider the above-results for plane wave normal modes to be applicable for the ``typical,'' or ``average'' case.

\begin{figure}
\includegraphics[width=3.25in]{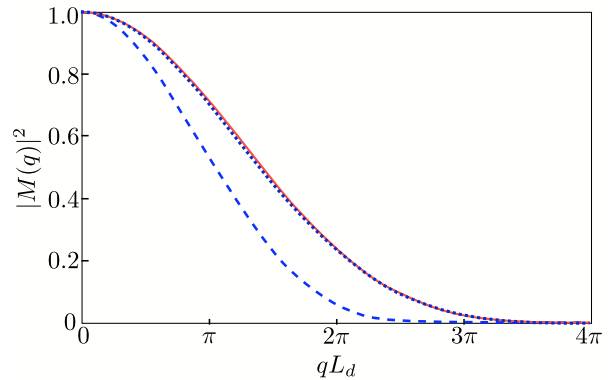}
 \caption[]
{(color online) Longitudinal form factor $|M(q)|^2$ for symmetric square well confinement as described in Appendix \ref{AppOverlap}, with barrier separation $L_d = 100\ {\rm nm}$, for $E_g = 0.03\ {\rm eV}$ (dashed blue) and $E_g = 1\ {\rm eV}$ (solid red).
For the large gap tube, $kL_d \approx \pi$ and $|M(q)|^2$ closely follows the hard wall limit, Eq.(\ref{MSupp}),
with strong suppression for $qL_d \gtrsim 2\pi$.
For the small gap tube, softer confinement results in a larger effective length $L_d^*$, $kL_d^* = \pi$.
When plotted against $qL_d^*$, the profile nearly collapses onto the large gap result (dotted line).
}
\label{FigMq}
\end{figure}
In section \ref{SpinPhononCoupling} we pointed out that for small momentum (energy) transfer $q\rightarrow 0$, the factor $M(q) \rightarrow 1$.
For large momentum transfer $qL_d \gg 2\pi$, however, $M(q)$ is strongly suppressed due to the fast oscillations of $e^{iqz}$ relative to the electron wavefunction.
Using the results of Appendix \ref{AppOverlap} for symmetric square well confinement, we plot $|M(q)|^2$ for a large gap and a small gap tube in Fig.\ref{FigMq}.
For the small gap tube, where $kL_d \approx 0.75$, confinement is softer and the electron wavefunction extends over a distance significantly larger than the barrier separation $L_d$.
Thus suppression sets in at $qL_d^* \gg 2\pi$, where $kL_d^* = \pi$.  

The asymptotic behavior of $M(q)$ for large $q$ contains two contributions.
The classically allowed region provides a contribution $M(q) \sim 1/q^3$ for $qL_d \gg 2\pi$, see Eq.(\ref{MSupp}).
In the classically forbidden region, the exponential tails of the wave function contribute a Lorentzian decay law $M(q) \sim 1/(k_c^2 + q^2)$ with a prefactor that tends to zero as $k_cL_d \rightarrow \infty$.
Power law decay arises from the steps in the potential at $z = \pm L_d/2$, but the suppression behavior is generic.
For a smooth confinement potential one expects exponential decay, with a characteristic scale $qL_d^* \approx 2\pi$.

The resonant enhancement of relaxation due to the flexon density of states singularity is thus not affected by $M(q)$ 
as long as the minimum of the energy gap $\epsilon(B_z,B_\perp) = \hbar\beta q^2$ at the upper avoided crossing corresponds to a phonon momentum $q$ satisfying $qL^*_d \lesssim 2\pi$.
Using $\hbar \beta \approx 10^4\ \mu{\rm eV (nm)}^2$ (see Sec.\ref{LowerKramers}) and an energy gap of $10\ \mu{\rm eV}$ as observed in Ref.[\onlinecite{Hugh}], we find $qL^*_d = 1.6 \mbox{--} 3.2$ for $L^*_d = 50 \mbox{--} 100\ {\rm nm}$.
Thus for the parameters of Ref.[\onlinecite{Hugh}], the longitudinal overlap does not suppress relaxation at the upper avoided crossing, but will lead to suppression for larger energy splittings.
Note that the same form factor appears in the matrix elements for the deformation potential coupling\cite{Bulaev} and for the g-factor anisotropy deflection coupling\cite{Borysenko}, and thus does not affect the relative competition between these mechanisms. 

In the presence of disorder, the quadratically dispersing flexons are subject to the same conditions that lead to the familiar localization of all one-dimensional electron eigenstates in a random potential\cite{Thouless, Berezinskii}.
For weak scattering characterized by $q\ell \gg 1$, where $\ell$ is the localization length, the density of states remains smooth and the normal mode profiles oscillate within envelopes with exponentially decaying tails, $e^{-|x|/\ell}$.
The exponential envelope cuts off the rapid oscillations responsible for the suppression of $M(q)$, allowing $\ell$ to substitute for $L_d$.
Thus localization can in some cases help promote relaxation at energy splittings larger than those where strong suppression by $M(q)$ sets in for clean samples. 
Because the localization length $\ell$ is of the order of the mean free path\cite{Thouless}, which for short range scatterers is proportional to the phonon energy, $\ell \propto q^2$ and the clean limit (very weak scattering) should generally be recovered for large enough energy transfers.

When scattering is strong, $q\ell \ll 1$, the normal mode spectrum collapses to a collection of discrete resonances, and the normal mode profiles are highly distorted.
In this case, the low energy modes responsible for the van Hove singularity that promotes spin relaxation near the upper avoided crossing can be destroyed.
Rather than observing a smooth increase of the relaxation rate as the intersection is approached, a complicated non-monotonic system of resonances may then be found.

It is also interesting to note that rigid rotational motion of the entire sample, as described in Fig.\ref{fig1}b, can be regarded as a global ($q = 0$) deflection of the nanotube. 
Consequently, simple laboratory vibrational noise can be a source of electron spin decoherence through the deflection coupling mechanism\cite{MarcusPC}.

In summary, we have identified and analyzed an efficient mechanism of electron spin relaxation in nanotube quantum dots that results from spin-orbit-mediated coupling between the electron spin and nanotube deflection.
Due to the flexon density of states singularity at small energies, relaxation due to this deflection coupling mechanism is particularly efficient near level crossings or small avoided crossings.
This mechanism is expected to dominate over other mechanisms such as deformation potential\cite{Bulaev} and anisotropic g-factor induced deflection coupling\cite{Borysenko}, which are suppressed by small phonon momentum $q$ and $g$-factor anisotropy, respectively.
Finally, we predict a robust minimum of the spin relaxation time $T_1$ near the upper avoided crossing in both the high and low temperature regimes, thus offering a firm basis for understanding the observed $T_1$ minimum of Ref.[\onlinecite{Hugh}].

{\it Note added in proof.} Recently, electrical spin manipulation in curved carbon nanotubes based on a related spin orbit coupling mechanism was proposed\cite{KarstenCharlie}.

\section{Acknowledgements}
We gratefully acknowledge H. O. H. Churchill, F. Kuemmeth, and C. M. Marcus for stimulating discussions.
The work of MR was supported by NSF grants PHY-0646094 and DMR-09-06475.
EIR received support from the NSF Materials World Network Program.

\appendix
\section{Longitudinal Form Factor $M(q)$}\label{AppOverlap}
As mentioned in section \ref{SpinPhononCoupling}, every matrix element $\MatEl{\psi_f}{H_{\rm s-ph}}{\psi_i}$ includes a longitudinal form factor $$M(q) = \int_{-\infty}^\infty dz\, n(z)\, e^{iqz},$$ 
which depends on the specific form of longitudinal confinement for the quantum dot, but is independent of the composition of $\Ket{\psi_{i(f)}}$ in terms of the basis states $\{\Ket{\tau\ s}\}$.
In this appendix we discuss the properties of $M(q)$ in more detail, and provide an explicit result for $M(q)$ in the case of square-well confinement.

The property that a unique longitudinal overlap integral $M(q)$ can be factored out of all matrix elements of $H_{\rm s-ph}$ independent of the initial and final states relies on two key facts.
First, the perturbation $H_{\rm s-ph}$, Eq.(\ref{HSPh}), is diagonal in the isospin index.
Thus for any initial and final states $\Ket{\psi_i}$ and $\Ket{\psi_f}$, which can be written as $\Ket{\psi_n} = \sum_{\tau,s} c_{n; \tau s}\Ket{\tau\ s}$, where $n$ stands for $i$ or $f$, 
\bea
\!\!\!\!\!\!\!\!\!\!\MatEl{\psi_f}{H_{\rm s-ph}}{\psi_i}\!\! &=&\!\!\! \sum_{\substack{ \tau, \tau^\prime\\ s,s^\prime}} c^*_{f;\tau s}c_{i; \tau's'}\MatEl{\tau\ s}{H_{\rm s-ph}}{\tau'\ s'}\nonumber\\
&=&\!\!\!\! \sum_{\substack{\tau, s,s'}}\!\! c^*_{f;\tau s}c_{i; \tau s'}\MatEl{\tau\ s}{H_{\rm s-ph}}{\tau\ s'}.
\eea
Furthermore, because each state $\Ket{\tau\ s} = \Ket{\tau}\otimes\Ket{s}$ is a product of orbital and spin parts, Eq.(\ref{HSPh}) gives
\be
\MatEl{\tau\ s}{H_{\rm s-ph}}{\tau\ s'} = \MatEl{\tau}{e^{iqz}}{\tau}\cdot \MatEl{s}{H_{{\rm s-ph}}^{\tau}}{s'},
\ee
where $H_{{\rm s-ph}}^{\tau}$ is the spin-phonon coupling Hamiltonian of Eq.(\ref{HSPh}) projected onto the orbital state $\Ket{\tau}$, with 
$$\MatEl{\tau}{e^{iqz}}{\tau} = \int_{-\infty}^{\infty}dz_0\, e^{iqz_0} \MatEl{\tau}{\delta(z - z_0)}{\tau}$$
factored out.
Second, 
the density $n(z_0) = \MatEl{\tau}{\delta(z - z_0)}{\tau}$, which includes a sum over densities on the two sublattices, is a real scalar.
Because $\Ket{\tau}$ with $\tau = 1$ and $\tau = -1$ are time-reversal conjugate, $n(z_0)$ is independent of $\tau$.
Hence 
\be
\MatEl{\psi_f}{H_{\rm s-ph}}{\psi_i} = M(q) \sum_{\tau,s,s'} c^*_{f;\tau s}c_{i; \tau s'} \MatEl{s}{H_{{\rm s-ph}}^{\tau}}{s'},
\ee
thus completing the proof that a single, unique contribution from the longitudinal degrees of freedom appears in all transition rates.

In the presence of a magnetic field that breaks time reversal symmetry, the density $n_{\tau,s}(z_0) = \MatEl{\tau\ s}{\delta(z - z_0)}{\tau\ s}$ is generally not independent of $\tau$ and $s$.
In particular, the term $\tau_3\mu_{orb}B_z$ in Eq.(\ref{H0}) leads to a renormalization of $E_g$ of opposite sign in valleys $K$ and $K'$.
For the device of Ref.[\onlinecite{Hugh}], with a gap $E_g \approx 30\ {\rm meV}$ and for fields up to $1.5\ {\rm T}$, however, $\mu_{orb}B/E_g \lesssim 1/30$.
Thus our calculation of the longitudinal wave functions to lowest order and the resulting factorization of $M(q)$ is justified.

\begin{figure}
\includegraphics[width=3.25in]{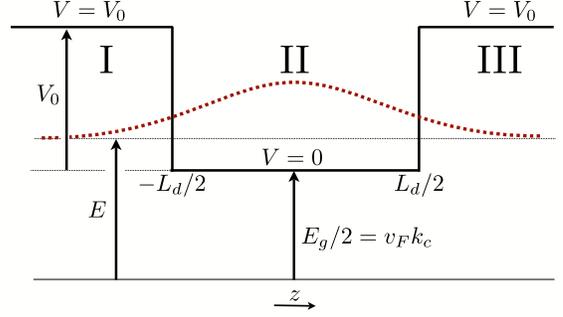}
\caption[]
{(color online) We consider a quantum dot formed in a (small or large gap) semiconducting nanotube.
The gate-induced potential $V(z)$ is taken to be: $V(z) = V_0$ for $z < -L_d/2$ or $z > L_d/2$, and $V(z) = 0$ for $-L_d/2 < z < L_d/2$.
The band gap $E_g$ arises either due to quantization of transverse motion for pure semiconducting tubes, in which case $E_g \propto 1/R$, or due to curvature of the graphene sheet, in which case $E_g \propto 1/R^2$. 
Electron density $n(z)$ is indicated by the red (dotted) curve.
}
\label{FigApp}
\end{figure}

We now calculate $M(q)$ for a quantum dot in a carbon nanotube formed by a symmetric step potential as 
shown in Fig.\ref{FigApp}.  
Although the details of $M(q)$ depend on the form of the longitudinal confinement, this example illustrates its basic properties.
Electron eigenfunctions in a square well potential were analyzed in detail in Ref.[\onlinecite{Bulaev}].
Therefore here we focus on the main features and omit a step-by-step derivation.

A pure ``hard-wall'' boundary condition cannot be defined for this system due to the relativistic dispersion and internal spinor structure (recall Klein phenomenon for relativistic particles in one dimension).
The 
strongest confinement is realized when the potential step at the barrier is roughly equal to half the band gap, $V_0 \approx E_g/2$.
Here the tails of the wave function decay exponentially over a distance proportional to $E_g^{-1}$.
For large gap tubes, 
$E_g \propto 1/R$, while for small gap tubes, the curvature-induced gap scales according to $E_g \propto \cos 3\eta/R^2$, where $R$ is the tube radius and $\eta$ is the winding angle.

The orbital states $\Ket{\tau}$ with $\tau = \pm 1$ of Eq.(\ref{H_d}) correspond to the lowest quantized mode of longitudinal motion in the $K$ and $K'$ valleys, respectively.
Because we assume that the external potential $V(z)$ 
only depends on the longitudinal coordinate $z$ and not on the circumferential coordinate $x$, these eigenfunctions of $H_d$ can be factored as
\be
\begin{array}{ccc}
 \Amp{\vec{r}}{\tau} \propto e^{i \vec{K}_\tau\cdot\vec{r}}\,e^{\tau i k_c x}\,\, \psi_\tau(z),\\
\end{array}
\ee
where $\vec{r} = (x, z)$, the circumferential wavevector $k_c$ is proportional to $E_g$, $E_g = 2\hbar v_Fk_c$, and $\psi_\tau$ is a two-component wavefunction describing amplitudes $\psi_\tau^{\rm A}(z)$ and $\psi_\tau^{\rm B}(z)$, on the graphene A and B sublattices.
Recall that $K_\tau = \tau 2\pi/3a(1, \sqrt{3})$ for $\tau = \pm 1$ corresponds to the $K$ ($\tau = 1$) or $K'$ ($\tau = -1$) point of the graphene Brillouin zone, where $a$ is the lattice constant.
In terms of the pseudospin Pauli matrices $\sigma_1$ and $\sigma_2$, $\psi_\tau$ satisfies
\be
\left[\hbar v_F\left(\tau\sigma_1k_c - i\sigma_2 \frac{d}{dz}\right) + V(z)\right]\,\psi_{\tau} = E\, \psi_{\tau},
\ee
see Eq.(\ref{H_d1}).
In a homogeneous tube with constant $V(z)$, the eigenstates are plane waves proportional to $e^{ikz}$ and the spectrum has a relativistic dispersion $E = \pm \hbar v_F\sqrt{k_c^2 + k^2}$. 


In the classically forbidden region I, $z < -L_d/2$, the potential takes the constant value $V(z) = V_0$.  
To emphasize the essential behavior and to simplify the equations, we pick the special value $V_0 = E$, where $E$ is determined self-consistently through the allowed wavevector $k$ which satisfies the continuity relations at the boundaries, see Eq.(\ref{kCond}) below.
With this choice, the wavefunction in the classically forbidden region is described by a single exponential living on only one sublattice: 
the spinor components $\psi^{\rm A}_{\tau,\rm I}$ and $\psi^{\rm B}_{\tau,\rm I}$ satisfy 
\bea
\psi_{\tau,{\rm I}}^{\rm A}(z) &=& \frac{A_\tau}{2}\left({1 - \tau}\right)e^{k_c (z + L_d/2)},\nonumber\\
\psi_{\tau, {\rm I}}^{\rm B}(z) &=& \frac{A_\tau}{2}\left({1 + \tau}\right)e^{k_c (z + L_d/2)}.
\eea
Similarly, in region III, $x > L_d/2$,
\bea
\psi_{\tau,{\rm III}}^{\rm A}(z) &=& \frac{D_\tau}{2}\left({1 + \tau}\right)e^{-k_c (z - L_d/2)},\nonumber\\
\psi_{\tau, {\rm III}}^{\rm B}(z) &=& \frac{D_\tau}{2}\left({1 - \tau}\right)e^{-k_c (z-L_d/2)}.
\eea
In these regions, the solutions in opposite valleys exist on opposite sublattices. 
However, the {\it total} density $n(|z|)$ is indeed the same for each case.

In the classically allowed region II, $V(z) = 0$ and the spinor components are plane waves $\psi^{\rm A}_{\tau, \rm II}\propto e^{ikz}\tau e^{\tau i\phi_k}$ 
and $\psi^{\rm B}_{\tau,\rm II}\propto e^{ikz}$ which satisfy:
\be
\hbar v_F\!\!\left(\begin{array}{cc} 0 & \tau k_c - ik\\\tau k_c + ik & 0\end{array}\right)\!\!
 \left(\!\!\begin{array}{c}\tau e^{i\tau\phi_k}\\ 1\end{array}\!\!\right)
=
E  \left(\!\!\begin{array}{c}\tau e^{i\tau\phi_k}\\ 1\end{array}\!\!\right).
\ee
Here $e^{i\phi_k} = (k_c - ik)/\sqrt{k_c^2 + k^2}$, and  $E = \hbar v_F\sqrt{k_c^2 + k^2}$.
Note that $\phi_{-k} = -\phi_k$.

Due to the degeneracy of states $k$ and $-k$, in region II a general state with energy $E$ has the form:
\be
\left(\begin{array}{c}\psi^{\rm A}_{\tau,\rm II}\\ \psi^{\rm B}_{\tau,\rm II}\end{array}\right) = B_\tau e^{ikz}\left(\begin{array}{c}\tau e^{i\tau\phi_k}\\ 1\end{array}\right) + C_\tau e^{-ikz}\left(\begin{array}{c}\tau e^{-i\tau\phi_k}\\ 1\end{array}\right).
\ee

Using the continuity condition of the wavefunction at the boundaries between different regions, we fix the relative coefficients $A_\tau, B_\tau, C_\tau$, and $D_\tau$, and find the allowed wavevector $k$ that gives a bound state~\footnote{For comparison, see Bulaev, Trauzettel, and Loss Eq.(11), with $V_g = E_{\kappa_m,k_n}$ and $\kappa_m \rightarrow k_c$.},
\be
\label{kCond}\tan k L_d = -\frac{k}{k_c},
\ee
which comes from the condition $kL_d = \phi_k \mod \pi$.

We calculate $M(q) = M_{\rm I}(q) + M_{\rm II}(q) + M_{\rm III}(q)$ piecewise.
In region I, $ M_{\rm I}(q) = \int_{-\infty}^{-z_*}dz\,|A_\tau|^2e^{2k_cz_*} e^{(2k_c + iq)z}$ evaluates to 
\bea
\label{MI} M_{\rm I}(q) 
&=&  \frac{4\mathcal{N}\sin^2 (kL_d)\, e^{-iqz_*}}{2k_c + iq},
\eea
where $z_* = L_d/2$ and the normalization constant $\mathcal{N}$ satisfies 
\be
\frac{1}{\mathcal{N}} = 4 L_d + \frac{4}{k_c}\sin^2 kL_d - \frac{2}{k}\sin 2kL_d. 
\ee
In region III, 
\be
 M_{\rm III}(q) = \frac{4\mathcal{N}\sin^2(kL_d)\,e^{iqz_*}}{2k_c - iq} = M_{\rm I}^*(q).
\ee
Finally, in region II we must evaluate
\be
M_{\rm II}(q) = 4\mathcal{N} \int_{-z_*}^{z_*}\!\!\! dz\left[\sin^2 k\!\left(z + z_*\right) + \sin^2 k\!\left(z - z_*\right)\right] e^{iqz},
\ee
which after some steps yields:
\begin{widetext}
\be
\label{MII} M_{\rm II}(q) = 4\mathcal{N}\left[\frac{2\sin qL_d/2}{q} - \cos kL_d \left(\frac{\sin (k + q/2)L_d}{2k + q} + \frac{\sin (k - q/2)L_d}{2k - q}\right)\right].
\ee
\end{widetext}
Because the density $n(z)$ is symmetric in $z$, 
$M(q) = M_{\rm I}(q) + M_{\rm II}(q) + M_{\rm III}(q)$ is real, and even in $q$.

For a high aspect ratio dot with $k_cL_d \gg 1$, we have $kL_d \approx \pi$.
In this case, the contributions $M_{\rm I}$ and $M_{\rm III}$ from the tails vanish, and Eq.(\ref{MII}) simplifies to
\bea
\label{MSupp} M_{\rm II}(q) = \frac{8\pi^2\sin(qL_d/2)}{qL_d[4\pi^2 - (qL_d)^2]},
\eea
which equals 1 for $q = 0$ and decays as $(qL_d)^{-3}$ for $qL_d \gg 2\pi$.
This limit coincides with Eq.(56) of Ref.[\onlinecite{Bulaev}].
Interestingly, for a small-gap nanotube with $E_g \approx 30\ {\rm meV}$ and $L_d = 100\ {\rm nm}$ as in Refs.[\onlinecite{Ferdinand,Hugh}], $kL_d \approx 0.75 \pi$,
indicating significant penetration into the classically forbidden region.


\begin{references}
\bibitem{nonadiab}
This cancellation is violated only when nonadiabatic corrections originating from the dependence of spin-phonon coupling on lattice velocities are taken into account. Their effect is small\cite{Borhani,SanJose} and we disregard it.
\bibitem{footnoteI}
While spin-orbit coupling is a generic feature of nanotubes, spin-split spectra such as those obtained in Refs.[\onlinecite{Ferdinand,Hugh}] and shown in Fig.\ref{fig1}a are not universally observed. The origin of this phenomenon, and in particular of the narrow Zeeman line of Ref.[\onlinecite{Chauvet}] with $g\approx 2$, is not currently understood.  It appears that the discrepancy might be attributed to motional narrowing of the spin-flip transitions due to strong scattering on imperfections of the nanotube\cite{MR}.
\end{references}
\end{document}